\documentstyle[12pt,epsfig]{article}
\newlength{\dinwidth}                       
\newlength{\dinmargin}                      
\def\cm{{\cal M}}
\def\bom#1{{\mbox{\boldmath $#1$}}}
\def\ep{\epsilon}
\def\beq{\begin{equation}}
\def\eeq{\end{equation}}
\def\beeq{\begin{eqnarray}}
\def\eeeq{\end{eqnarray}}
\def\pl#1#2#3{Phys.\ Lett.\ #1B (19#3) #2}

\def\np#1#2#3{Nucl.\ Phys.\ B#1 (19#3) #2}
\def\pr#1#2#3{Phys.\ Rev.\ D #1 (19#3) #2}
\begin{document}

\begin{titlepage}
\renewcommand{\thefootnote}{\fnsymbol{footnote}}
\begin{flushright}
     CERN-TH/96-240 \\ hep-ph/9609521
     \end{flushright}
\par \vspace{10mm}
\begin{center}
{\Large \bf
NLO QCD calculations in DIS at HERA based on the dipole
formalism\footnote{Presented by 
M.H. Seymour at the Workshop on Future Physics at HERA, DESY, Hamburg, Germany.
To appear in the Proceedings.}}
\end{center}
\par \vspace{2mm}
\begin{center}
{\bf S. Catani}\\

\vspace{5mm}

{I.N.F.N., Sezione di Firenze}\\
{and Dipartimento
di Fisica, Universit\`a di Firenze}\\
{Largo E. Fermi 2, I-50125 Florence, Italy}

\vspace{5mm}

{\bf M.H. Seymour}\\

\vspace{5mm}

{Theory Division, CERN}\\
{CH-1211 Geneva 23, Switzerland}
\end{center}

\par \vspace{2mm}
\begin{center} {\large \bf Abstract} \end{center}
\begin{quote}
We briefly describe a new general algorithm for carrying out QCD
calculations to next-to-leading order in perturbation theory.
The algorithm can be used for computing arbitrary jet cross sections in
arbitrary processes and can be straightforwardly implemented in 
general-purpose Monte Carlo programs.
We show numerical results for the specific case of jet cross sections
in deep inelastic scattering at HERA energies.
\end{quote}
\vspace*{\fill}
\begin{flushleft}
     CERN-TH/96-240 \\ September 1996
\end{flushleft}
\end{titlepage}

\vspace*{1cm}
\begin{center}  \begin{Large} \begin{bf}
NLO QCD calculations in DIS at HERA based on the 
  dipole formalism\\
  \end{bf}  \end{Large}
  \vspace*{5mm}
  \begin{large}
S. Catani$^a$, M.H. Seymour$^{b}$\\
  \end{large}
\end{center}
$^a$ I.N.F.N., Sezione di Firenze,
        and Dipartimento di Fisica, Universit\`a di Firenze, \\
        Largo E. Fermi 2, I-50125 Florence, Italy\\
$^b$ Theory Division, CERN,
        CH-1211 Geneva 23, Switzerland\\
\begin{quotation}
\noindent
{\bf Abstract:}
We briefly describe a new general algorithm for carrying out QCD
calculations to next-to-leading order in perturbation theory.
The algorithm can be used for computing arbitrary jet cross sections in
arbitrary processes and can be straightforwardly implemented in 
general-purpose Monte Carlo programs.
We show numerical results for the specific case of jet cross sections
in deep inelastic scattering at HERA energies.
\end{quotation}
\section{Introduction}
\vspace{-1.5ex}

In order to make quantitative predictions in perturbative QCD, it is
essential to work to (at least) next-to-leading order (NLO).  However, this
is far from straightforward because for all but the simplest quantities,
the necessary phase-space integrals are too difficult to do analytically,
making numerical methods essential.  But the individual integrals are
divergent, and only after they have been regularized and combined is the
result finite.  The usual prescription, dimensional regularization,
involves working in a fractional number of dimensions, making analytical
methods essential.

To avoid this dilemma, one must somehow set up the calculation such that
the singular parts can be treated analytically, while the full complexity
of the integrals can be treated numerically.  Efficient techniques have
been set up to do this, at least to NLO, during the last few years.

A new general algorithm was recently presented~\cite{CS}, which can be used
to compute arbitrary jet cross sections in arbitrary processes.  It is
based on two key ingredients: the {\em subtraction method\/} for cancelling
the divergences between different contributions; and the {\em dipole
  factorization theorems} 
for the universal (process-independent) analytical 
treatment of individual divergent terms.  
These are sufficient to write a general-purpose
Monte Carlo program in which any jet quantity can be calculated simply by
making the appropriate histogram in a user routine.

In this contribution we give a brief summary of these two ingredients (more
details and references to other general methods can be found in
Refs.~\cite{CS}--\cite{CSrh}) and show numerical results for the
specific case of jets in deep-inelastic lepton-hadron scattering (DIS) at HERA
energies.

A Monte Carlo program for jet calculations in DIS 
based on a different method~\cite{GGK} is presented in Ref.~\cite{mepjet}.
Previous calculations~\cite{projet,disjet} were limited to a specific jet 
definition and used approximated methods.

\vspace{-1ex}
\section{The Dipole Formalism}
\vspace{-1.5ex}

\subsection{The subtraction method}
\vspace{-1.5ex}


Considering the case of DIS (full details for other processes are given in 
Ref.~\cite{CS}), 
the general structure of a QCD cross section in NLO is:
\beq
\label{shad}
\sigma = \sum_a f_a \times ( \sigma_a^{LO} + \sigma_a^{NLO} ) \;\;
\eeq
where the symbol $\times$ denotes the convolution with the  
density $f_a$ of the parton $a$ in the incoming hadron and 
the leading-order (LO) cross section $\sigma_a^{LO}$ is obtained by 
integrating the fully
exclusive Born cross section $d\sigma_a^{B}$ over the phase space for the
corresponding jet quantity. We suppose that this LO calculation involves
$m$ final-state partons, and write:
\beq
\label{sLO}
\sigma_a^{LO} = \int_m d\sigma_a^{B} \;.
\eeq
At NLO, we receive contributions from real and virtual processes (we assume
that the ultraviolet divergences of the virtual term are already
renormalized) and from a collinear counterterm that is necessary 
to define the scale-dependent parton densities:
\beq
\label{sNLO}
\sigma_a^{NLO} 
= \int_{m+1} d\sigma_a^{R} + 
\int_{m} \left[ d\sigma_a^{V} + d\sigma_a^{C} \right]\;.
\eeq
As is well known, each of these is separately divergent, although their sum
is finite.  These divergences are regulated by working in $d=4-2\epsilon$
dimensions, where they are replaced by singularities in $1/\epsilon$.  Their
cancellation only becomes manifest once the separate phase space integrals
have been performed.

The essence of the subtraction method~\cite{subm} is to use 
the {\em exact\/} identity
\beq
\label{sNLO1}
\sigma_a^{NLO} = \int_{m+1} \left[ d\sigma_a^{R} - d\sigma_a^{A}  \right] 
+ \int_{m+1} d\sigma_a^{A}
+  \int_m \left[ d\sigma_a^{V} +  d\sigma_a^{C} \right] \;,
\eeq
which is obtained by subtracting and adding back the `approximate' (or
`fake') cross section contribution $d\sigma_a^{A}$, which has to fulfil two
main properties.
Firstly, it must exactly match the singular behaviour (in $d$ dimensions)
of $d\sigma_a^{R}$ itself. Thus it acts as a {\em local\/} counterterm for
$d\sigma_a^{R}$ and one can safely perform the limit $\ep \to 0$ under the
integral sign in the first term on the right-hand side of
Eq.~(\ref{sNLO1}).
Secondly, $d\sigma_a^{A}$ must be analytically integrable (in $d$ dimensions)
over the one-parton subspace leading to the 
divergences. Thus we can rewrite the integral in the last term of
Eq.~(\ref{sNLO1}), to obtain
\beq
\label{sNLO2}
\sigma_a^{NLO} = \int_{m+1} \left[ \left( d\sigma_a^{R} \right)_{\ep=0}
- \left( d\sigma_a^{A} \right)_{\ep=0}  \;\right] +
\int_m 
\left[ d\sigma_a^{V} +  d\sigma_a^{C} + \int_1 d\sigma_a^{A} 
\right]_{\ep=0}\;\;.
\eeq
Performing the analytic integration $\int_1 d\sigma_a^{A}$, one obtains 
$\ep$-pole
contributions that can be combined with those in $d\sigma_a^{V}$ and
$d\sigma_a^{C}$, thus
cancelling all the divergences.  Equation~(\ref{sNLO2}) can be easily
implemented in a `partonic Monte Carlo' program that generates
appropriately weighted partonic events with $m+1$ final-state partons and
events with $m$ partons.

\vspace{-1ex}
\subsection{Dipole factorization and universal subtraction term}
\vspace{-1.5ex}

The fake cross section 
$d\sigma_a^{A}$ can be constructed in a
fully process-independent way, by using the factorizing properties of gauge
theories.  Specifically, in the soft and collinear limits, which give rise
to the divergences,
the factorization theorems can be used to write the
cross section as the contraction of the Born cross section with universal
soft and collinear factors (provided that colour and spin correlations are
retained).  However, these theorems are only valid in the exactly singular
limits, and great care should be used in extrapolating them away from these
limits.
In particular, a careful treatment of momentum conservation is
required. Care has also to be
taken in order to avoid double counting the soft and collinear divergences in
their overlapping region (e.g.~when a gluon is both soft and collinear to
another parton).
The use of the dipole factorization theorem introduced in Ref.~\cite{CSlett}
allows one to overcome these difficulties in a straightforward way.

The dipole factorization formulae 
relate the singular behaviour of $\cm_{m+1}$, the tree-level matrix element
with $m+1$ partons, to $\cm_{m}$. They
have the following symbolic structure:
\beq
\label{Vsim}
|\cm_{m+1}(p_1,...,p_{m+1})|^2 =
|\cm_{m}({\widetilde p}_1,...,{\widetilde p}_{m})|^2 
\otimes {\bom V}_{ij}
+ \dots \;\;.
\eeq
The dots on the right-hand side stand for contributions that are not singular 
when $p_i\cdot p_j \to 0$. 
The dipole splitting functions ${\bom V}_{ij}$ are universal 
(process-independent) singular factors that
depend on the momenta and quantum numbers of the $m$ partons in the tree-level
matrix element $|\cm_{m}|^2$. Colour and helicity correlations are denoted by
the symbol $\otimes$. The set ${\widetilde p}_1,...,{\widetilde p}_{m}$
of modified  momenta on the right-hand side of Eq.~(\ref{Vsim})
is defined starting from the original $m+1$ parton momenta in such a way that
the $m$ partons in $|\cm_{m}|^2$ are physical, that is, 
they are on-shell and energy-momentum conservation is
implemented exactly.
The detailed expressions for these parton momenta and for the dipole splitting
functions are given in Ref.~\cite{CS}.

Equation~(\ref{Vsim}) provides a {\em single\/} formula that
approximates the real matrix element $|\cm_{m+1}|^2$
for an arbitrary process, in {\em all\/} of its singular limits. These limits
are approached smoothly, avoiding double counting
of overlapping soft and collinear singularities.  Furthermore, the precise
definition of the $m$ modified
momenta allows an {\em exact\/} factorization
of the $m+1$-parton phase space, so that the universal dipole splitting
function can be integrated once and for~all.

This factorization, which is valid for the total phase space,
is not sufficient to provide a universal fake cross
section however, as its phase space
should depend 
on the particular jet
observable being considered.  The fact that the $m$ parton momenta are
physical provides a simple way to implement this dependence.  
We construct $d\sigma_a^{A}$ by adding the dipole contributions on the 
right-hand side of Eq.~(\ref{Vsim}) and for
each 
contribution 
we calculate the jet observable not
from the original $m+1$ parton momenta, but from the corresponding $m$
parton momenta, ${\widetilde p}_1,...,{\widetilde p}_{m}$.  Since these are
fixed during the analytical integration, it can be performed without any
knowledge of the jet observable.

\vspace{-1ex}
\section{Final Results}
\vspace{-1.5ex}

Refering to Eq.~(\ref{sNLO2}), the final procedure is then straightforward.
The calculation of any jet quantity to NLO consists of an $m+1$-parton
integral and an $m$-parton integral.  These can be performed separately
using standard Monte Carlo methods.

For the $m+1$-parton integral, a phase-space point is generated and the
corresponding real matrix element
in $d\sigma_a^R$ is
calculated.  These are passed to a user
routine, which can analyse the event in any way and histogram any
quantities of interest.  Next, for each dipole term 
(there are at most 10 of them in the calculation of (2+1)-jet 
observables in DIS)
in $d\sigma_a^A$, 
the set of $m$ parton momenta is derived from
the same phase-space point
and the corresponding dipole contribution is
calculated.  These are also given to the user routine.  They are such that
for any singular $m+1$-parton configuration, one or more of the $m$-parton
configurations becomes indistinguishable from it, so that they fall in the
same bin of any histogram.  Simultaneously, the real  
matrix element
and dipole term will have equal and opposite weights, so that the total
contribution to that histogram bin is finite.  Thus the first integral of
Eq.~(\ref{sNLO2}) is finite.

The $m$-parton integral 
in Eq.~(\ref{sNLO2})
has a simpler structure: it is identical
to the LO integration in Eq.~(\ref{sLO}), but with the Born term
replaced by the finite sum of the virtual matrix element 
in $d\sigma_a^V$, the collinear counterterm $d\sigma_a^C$
and the analytical integral of the dipole contributions
in $d\sigma_a^A$ (to be precise, the second term on the right-hand side
of Eq.~(\ref{sNLO2}) involves an additional one-dimensional 
convolution~\cite{CS}, which is a finite remainder that is left after the 
cancellation of the singularities in $d\sigma_a^C$).

Note that our algorithm does not require the convolution
with the parton densities $f_a$ to be made during Monte Carlo
integration.  One is free to choose either to calculate the hadron-level
cross section in Eq.~(\ref{shad}), thus including the convolution, 
or the parton-level cross section $\sigma_a\ = \sigma_a^{LO} + \sigma_a^{NLO}$ 
as a function of the partonic momentum fraction.  The latter can then be
convoluted with the parton densities after Monte Carlo integration.
This can be extremely useful in many respects. For instance, 
one can produce cross sections
with a wide
variety of parton densities, or study the scheme- and
scale-dependence of the results without having to reintegrate for each new
scheme or scale. 

\begin{figure}
  \centerline{\hfill\epsfig{figure=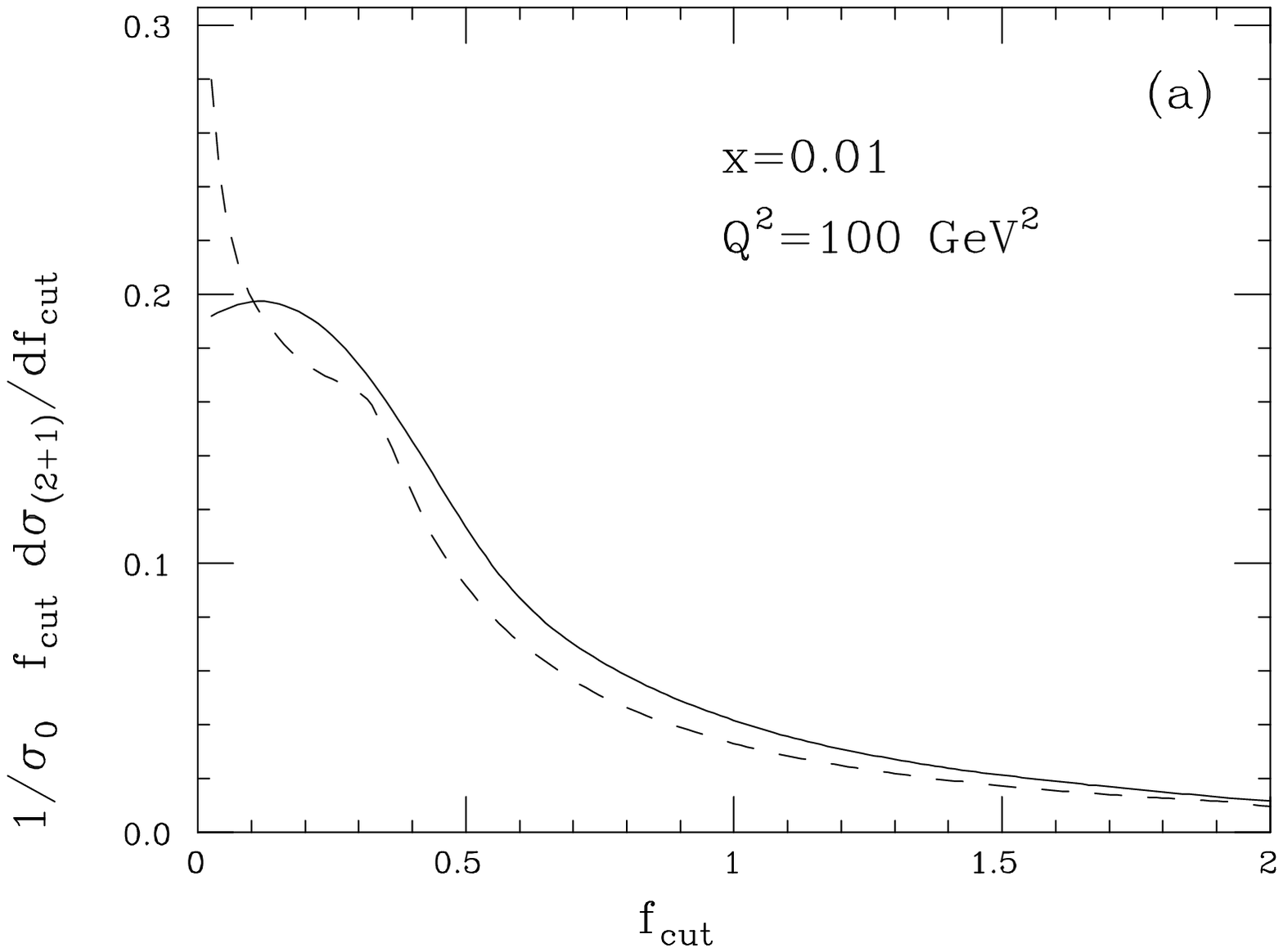,height=4.5cm}\hfill
              \hfill\epsfig{figure=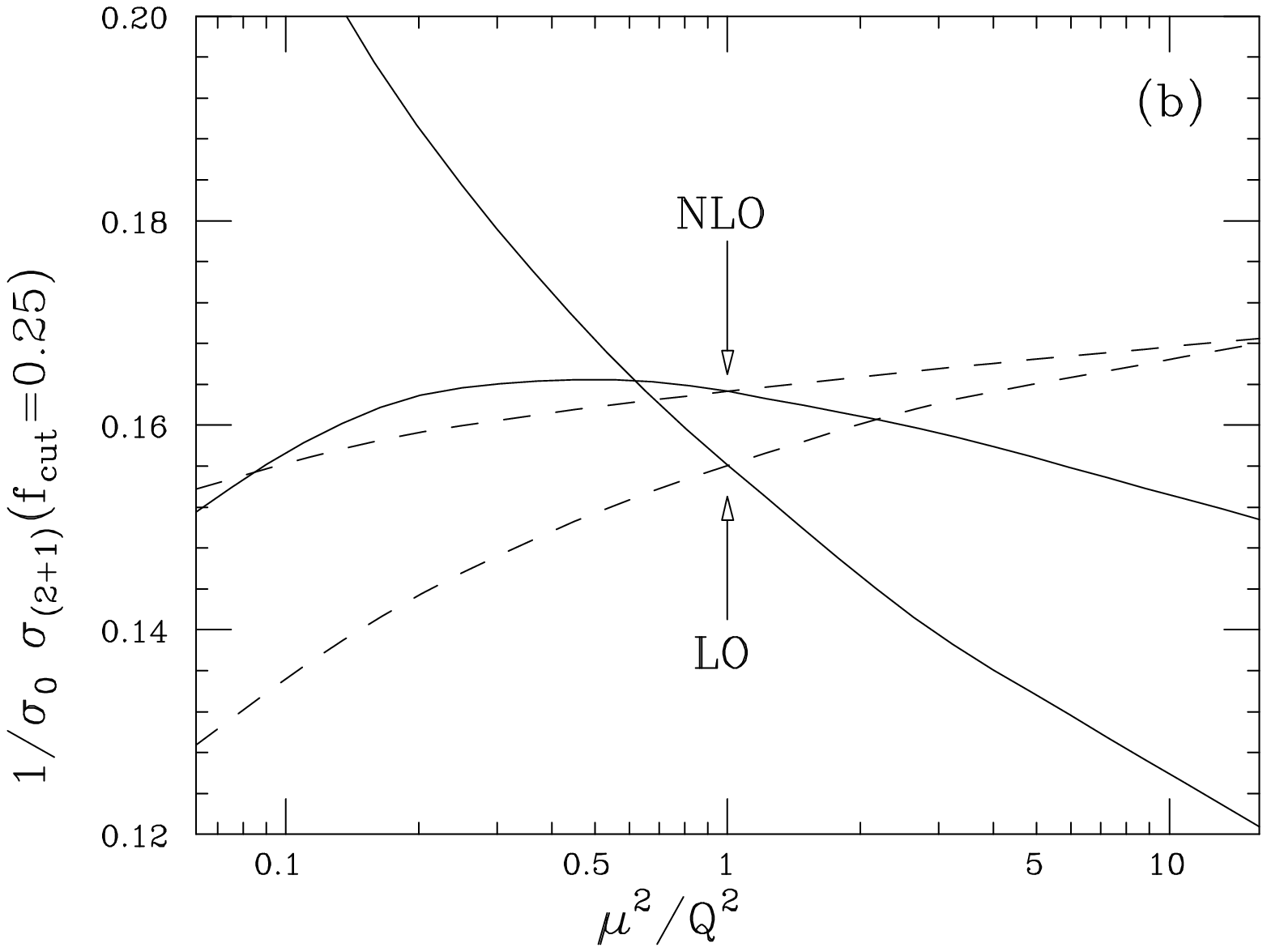,height=4.5cm}\hfill}
\caption[]{ Jet cross sections in $ep$ collisions at HERA energies 
            (${\sqrt s}= 300~{\rm GeV}$).
           (a) The distribution of resolution parameter $f_{cut}$ at which
                DIS events are resolved into $(2+1)$ jets according to the
                $k_\perp$ jet algorithm.  Curves are LO (dashed) and NLO
                (solid) using factorization and renormalization scales
                equal to $Q^2$, and the MRS D$-'$ distribution functions.
                Both curves are normalized to the LO cross section.
           (b) The rate of events with exactly $(2+1)$ jets at
                $f_{cut}=0.25$ with variation of renormalization
                (solid) and factorization (dashed) scales.  Normalization
                is again the LO cross section with fixed factorization
                scale.
\label{fig}}
\end{figure}

For the specific case of jet observables in DIS, we have
implemented the algorithm as a Monte Carlo program,
which can be obtained
from 
\verb+http://surya11.cern.ch/users/seymour/nlo/+. The program uses the matrix
elements evaluated by the Leiden group~\cite{leiden}. In Fig.~\ref{fig}a
we show as an example the differential jet rate as a function of jet
resolution parameter, $f_{cut}$,
using the $k_\perp$ jet algorithm~\cite{ktalg}.  We see that the NLO 
corrections are
generally small and positive, except at very small
$f_{cut}$
(where large logarithmic terms, $-\alpha_s\log^2f_{cut}$ arise at each
higher order).
In Fig.~\ref{fig}b, we show the variation
of the  
jet rate at a fixed $f_{cut}$ with factorization and
renormalization scales. The scale dependence is considerably smaller at NLO.
 
\vspace{-1ex}
\section{Conclusion}
\vspace{-1.5ex}

The subtraction method provides an {\em exact\/} way to calculate arbitrary
quantities in a given process using a general purpose Monte Carlo program.
The dipole formalism provides a way to construct such a program from
process-independent components.  Recent applications have included jets in
DIS.  

We have constructed a Monte Carlo program that can be used to carry out
NLO QCD calculations for any infrared- and collinear-safe observable 
(jet cross sections using different jet algorithms, event shapes, energy 
correlations and so forth) in $(2+1)$-jet configurations in DIS.
Possible applications to future physics at HERA include determinations of the
strong coupling constant $\alpha_S(Q)$ and extraction of the parton densities.
More details of the program, and its results, will be given
elsewhere.

\vspace{0.2cm}
\noindent{\bf Acknowledgments.}
This research is supported in part by EEC Programme 
{\it Human Capital and Mobility}, Network {\it Physics at High
Energy Colliders}, contract CHRX-CT93-0357 (DG 12 COMA). We wish to thank
P.J.~Rijken and W.L.~van Neerven for having provided the matrix elements 
for us.


\begin{thebibliography}{99}

\bibitem{CS}
S.\ Catani and M.H.\ Seymour, preprint CERN-TH/96-29 (hep-ph/9605323).

\bibitem{CSlett}
S.\ Catani and M.H.\ Seymour, \pl{378}{287}{96}.

\bibitem{CSrh}
S.\ Catani and M.H.\ Seymour, preprint CERN-TH/96-181 (hep-ph/9607318),
preprint CERN-TH/96-239 (hep-ph/9609237).

\bibitem{GGK}
W.T.\ Giele, E.W.N.\ Glover and D.A.\ Kosower, \np{403}{633}{93}.
 
\bibitem{mepjet}
E.\ Mirkes and D.\ Zeppenfeld, \pl{380}{205}{96}, preprint TTP-96-30
(hep-ph/9608201) and in these Proceedings.

\bibitem{projet}
D.\ Graudenz,  \pr{49}{3291}{94}, Comp. Phys. Commun. 92 (1995) 65. 

\bibitem{disjet}
T.\ Brodkorb and E.\ Mirkes, preprint MAD-PH-821 (hep-ph/9404287).  

\bibitem{subm}
R.K.\ Ellis, D.A.\ Ross and A.E.\ Terrano, \np{178}{421}{81};
Z.\ Kunszt and P.\ Nason, in `Z Physics at LEP 1', CERN 89-08, vol.~1, p.~373;
Z.\ Kunszt and D.E.\ Soper, \pr{46}{192}{92}.

\bibitem{leiden}
E.B.\ Zijlstra  and W.L.\ van Neerven , \np{383}{525}{92}.

\bibitem{ktalg}
S.\ Catani, Yu.L.\ Dokshitzer and B.R.\ Webber, \pl{285}{291}{92},
\pl{322}{263}{94};
B.R.\ Webber, J. Phys. G19 (1993) 1567;
S.\ Catani, Yu.L.\ Dokshitzer, M.H.\ Seymour and B.R.\ Webber, 
\np{406}{187}{93}.

\end{thebibliography}
\end{document}